\documentclass{acmsiggraph}                     

\usepackage{booktabs} 
\usepackage{microtype}
\usepackage{amsmath}
\usepackage{amssymb}
\usepackage{units}
\DeclareMathAlphabet{\mathcal}{OMS}{cmsy}{m}{n}  
\usepackage{xfrac}
\usepackage[normalem]{ulem}
\usepackage{graphicx}  		
\usepackage{float}
\usepackage{url}
\usepackage{xspace}
\usepackage{color}
\usepackage[normalem]{ulem}  
\usepackage{enumitem}  		
\usepackage{tikz}			
\usepackage{calc}			
\usepackage{collcell}
\usepackage{tabularx}

\usepackage{multirow}
\usepackage{amsmath}
\usepackage{amssymb}
\usepackage{amsfonts}
\usepackage{eucal}
\usepackage[latin1]{inputenc}
\usepackage[normalem]{ulem}
\usepackage{listings}
\usepackage{mathrsfs}
\usepackage{subcaption}
\usepackage{cancel}
\usepackage[normalem]{ulem}
\usepackage{cancel}
\usepackage{tcolorbox}
\usepackage{verbatimbox}

\usepackage{graphicx}
\usepackage{color}
\usepackage{wrapfig}
\usepackage{ifthen}
\usepackage[]{algorithm2e}

\title{Notes on optimal approximations for importance sampling}

\author{
	Jacopo Pantaleoni\thanks{e-mail: jpantaleoni@nvidia.com}\\NVIDIA
	\\
	\\
	Eric Heitz\thanks{e-mail: eheitz.research@gmail.com}\\Unity Technologies
}


\begin{document}
	
	\newcommand{\picresdir}{final}

	\teaser{
		\includegraphics[width=100.0mm]{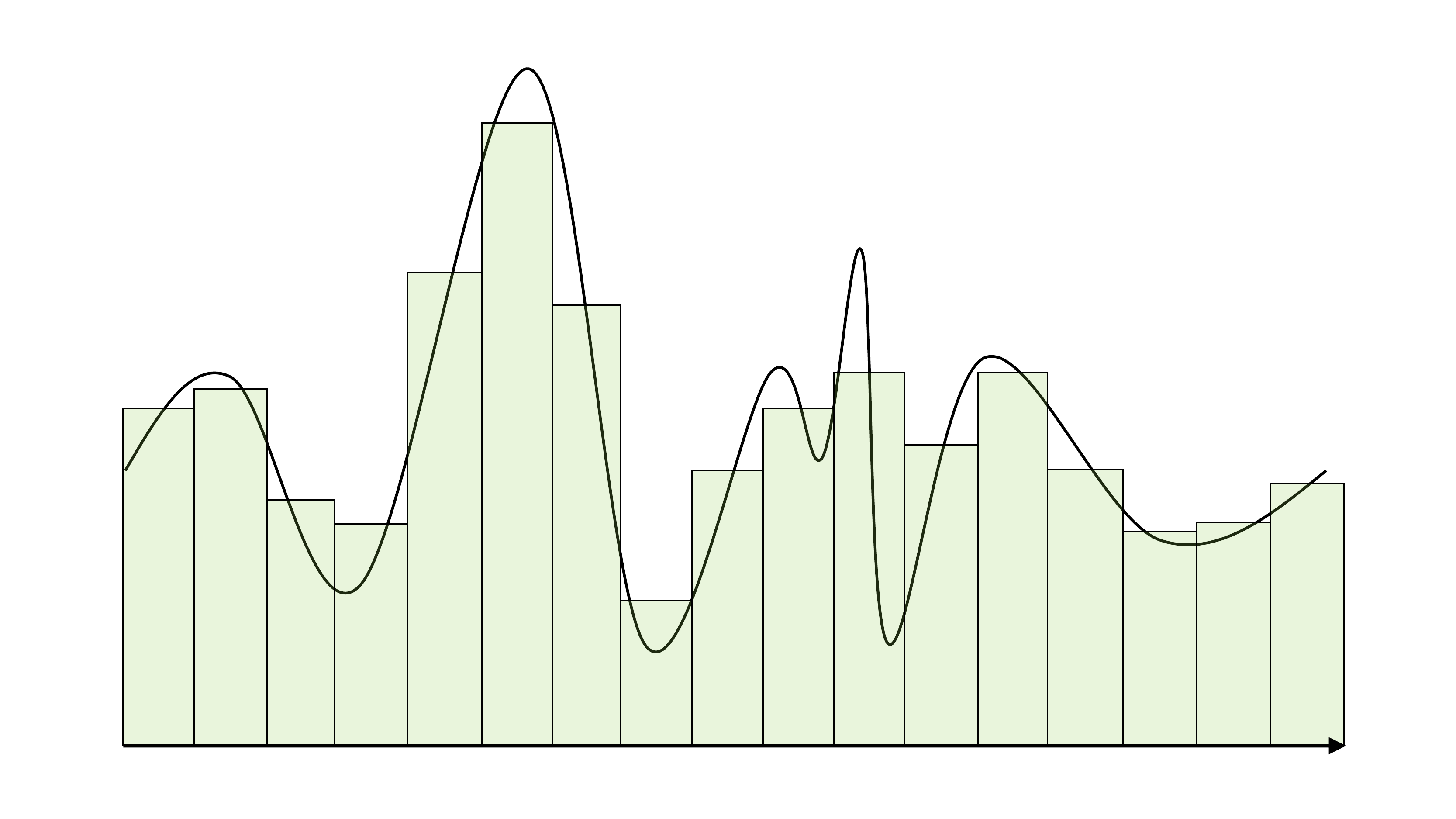}
		\caption{
			We seek to find projections onto elementary basis functions that result in optimal variance bounds when the projection is used as an importance sampling variate.
		}
	}
	
	
	\maketitle
	
	
	\begin{abstract}
	
		In this manuscript, we derive optimal conditions for building function approximations that minimize variance when used as importance sampling
	 	estimators for Monte Carlo integration problems.
	 	Particularly, we study the problem of finding the optimal projection $g$ of an
	 	integrand $f$ onto certain classes of piecewise constant functions, in order to minimize the variance of the unbiased importance sampling estimator $E_g[f/g]$, as well as the related problem of finding optimal mixture weights to approximate and importance sample a target mixture distribution $f = \sum_i \alpha_i f_i$ with components $f_i$ in a family $\mathcal{F}$, through a corresponding mixture of importance sampling densities $g_i$ that are only approximately proportional to $f_i$.
	 	We further show that in both cases the optimal projection is different from the commonly used $\ell_1$ projection, and provide an intuitive explanation for the difference.
	 	
	\end{abstract}

	

	\keywordlist
	
	
	\copyrightspace

\section{Introduction}

Importance sampling is one of the most important variance reduction tools in Monte Carlo integration.
Given a measurable space $(\Omega,\mu)$, and a function $f:\Omega \rightarrow {\mathbb R}$, suppose we want to estimate an integral of the form:
\begin{equation}
I = \int_\Omega f(x) d\mu(x) 
\end{equation}
Basic Monte Carlo integration tells us that we can approximate $\mu$ with the unbiased estimator:
\begin{equation}
I \approx \frac{1}{N} \sum_{j=1}^N f(X_j)
\end{equation}
where $X_j$ are realizations of a uniform random variable.
The most basic form of importance sampling states that if $X$ is a random variable with density $g$, the same integral can be equivalently calculated as:
\begin{equation}
I = E_g(f(X))
\end{equation}
and that one can hence construct the following Monte Carlo estimator:
\begin{equation}
I \approx F_N = \frac{1}{N} \sum_{j=1}^N \frac{ f(X_j) }{ g(X_j) }.
\end{equation}
The reason why importance sampling is interesting is that the variance of the above formula depends on how well $g$ approximates $f$:
\begin{equation}
V[F_N] = \frac{1}{N} \cdot V[f;g]
\end{equation}
with:
\begin{equation}
V[f;g] = \int_\Omega \frac{ f^2(x) }{ g(x) } d\mu(x) - I^2.
\end{equation}
In particular, if it was possible to simulate random variables from a distribution exactly proportional to $f$, i.e. with $g = f / I$, one would get a zero variance estimator:
\begin{eqnarray}
V[f;g] &=& \int_\Omega I \cdot \frac{ f^2(x) }{ f(x) } d\mu(x) - I^2 \nonumber \\
&=& I \cdot \int_\Omega f(x) d\mu(x) - I^2 = 0
\end{eqnarray}

Unfortunately, building such a zero variance estimator is most of the times not practical, as it would require knowledge of the integral we want to estimate in order to normalize the distribution $g$.
Hence, many algorithms rely on building useful approximations to $f$.

As a first application of this manuscript, we seek optimality conditions for such approximations, with a particular focus on the class of piecewise constant functions.

Since many such approximations are built as combinations of a finite set of basis functions $B = \left( b_1, \dots, b_n \right)$, we formulate the problem as that of finding a projection $g = P_B(f)$ that is optimal in the sense of minimizing variance from equation (6).
As a result, we will show that approximations resulting from the rather customary $\ell_1$ projection of $f$ onto $B$ does not lead to optimal estimators.

As a second application, we look at the related problem of estimating the integral of a mixture $f = \sum_i c_i f_i$ when the individual terms $f_i$ can only be approximately sampled, that is to say when we can only draw samples from densities $g_i \approx f_i$.
As a matter of fact, this latter application includes the first as a special case: it's sufficient to identify the basis functions $b_i$ with the densities $g_i$, and consider $f_i$ as the restrictions of $f$ to the support of $b_i$ - yet we thought it is interesting to consider both points of view separately, particularly as they lead to alternative solutions and different generalizations.

\section{Optimal Piecewise Constant Approximations}

Given a function $f:\Omega \rightarrow {\mathbb R}$, and a finite set of basis functions $B = \{b_i:\Omega \rightarrow {\mathbb R}\}_{i=1,...,n}$, we seek to find a projection $g = P_{is}(f)$:
\begin{equation}
g = \sum_i \alpha_i b_i(x)
\end{equation}
with the normalization constraint:
\begin{equation}
\sum_i \alpha_i = 1
\end{equation}
which minimizes the variance of the importance sampling estimator:
\begin{equation}
\mu = E[f/g] \approx \frac{1}{N} \sum_{j=1}^N \frac{f(X_j)}{g(X_j)}.
\end{equation}
That is, we want to minimize the following quantity:
\begin{equation}
V[f;g] = \int_\Omega \frac{ f^2(x) }{ g(x) } dx - \mu^2.
\label{BasicVar}
\end{equation}
Particularly, since we want to focus our analysis on piecewise constant functions $g$, we will look at elementary basis functions $b_i$ with non-overlapping compact supports $C_i$:
\begin{equation}
b_i(x) = |C_i|^{-1} \cdot \mathcal{X}_{C_i}(x) \nonumber
\end{equation}
with $C_i \cap C_j = \emptyset$ for $i \neq j$.
\vspace{3mm} \\
Regular $\ell_1$ projection would set the coefficients $\alpha_i$ as:
\begin{equation}
\alpha_i = \frac{1}{\mu} \int_\Omega f(x) b_i(x) dx = \frac{1}{\mu} \int_{C_i} f(x) dx;
\end{equation}
while this is commonly used, we here prove it is not the optimal choice.

\vspace{3mm}
\subsection{Solution}

Since $b_i$ have non overlapping supports, we can expand equation (\ref{BasicVar}) into:
\begin{equation}
V[f;g] = \sum_{i=1}^n \int_{C_i} \frac{ f^2(x) }{ \alpha_i b_i(x) } dx - \mu^2
\label{SimpVar}
\end{equation}
and if we define the constants:
\begin{equation}
m_i^{(2)} = \int_{C_i} \frac{ f^2(x) }{ b_i(x) } dx \nonumber
\end{equation}
we can further rewrite equation (\ref{SimpVar}) as:
\begin{equation}
V[f;g] = \sum_{i=1}^n \frac{ m_i^{(2)} }{ \alpha_i } - \mu^2.
\end{equation}
In order to find the minimum of $V[f;g]$, we use the method of Lagrange multipliers:
\begin{equation}
\nabla_{\alpha,\lambda} \left[ \sum_{i=1}^n \frac{ m_i^{(2)} }{ \alpha_i } - \mu^2 + \lambda \left(\sum_{i=1}^n \alpha_i - 1\right) \right] = 0
\end{equation}
which in turn translates into the set of $n$ equations:
\begin{eqnarray}
&& \frac{\partial}{\partial_{\alpha_i}} \left[ \frac{ m_i^{(2)} }{ \alpha_i } - \mu^2 + \lambda \alpha_i \right] = 0 \nonumber \\
&\Leftrightarrow& -\frac{ m_i^{(2)} }{ \alpha_i^2 } + \lambda = 0 \nonumber \\
&\Leftrightarrow& \alpha_i^2 = \frac{m_i^{(2)}}{\lambda} \nonumber \\
&\Leftrightarrow& \alpha_i = \sqrt{ \frac{m_i^{(2)}}{\lambda} }
\end{eqnarray}
subject to the constraint:
\begin{eqnarray}
&& \sum_i \sqrt{\frac{m_i^{(2)}}{\lambda}} = 1 \nonumber \\
&\Leftrightarrow& \lambda = \left(\sum_i \sqrt{m_i^{(2)}}\right)^2
\end{eqnarray}
leading to the final result:
\begin{equation}
\alpha_i =
\frac{
	\sqrt{m_i^{(2)}}
}{
	\sum_j \sqrt{m_j^{(2)}}
}.
\end{equation}
Notice how $P_{is}(f)$ is essentially a normalized $\ell_2$-norm projection, and may differ significantly from the more commonly used normalized $\ell_1$ projection.
While this result may be surprising at first, we believe it has a quite intuitive explanation: the optimal projection simply puts more weight on high values of the integrand, and does so proportionately to its square in order to minimize the quadratic variance functional.

\vspace{3mm}
\subsection{Generalization to arbitrary basis functions}
In the previous section we only considered basis functions with non-overlapping compact supports.
If we left this restriction out the original optimization problem would look like this instead:
\begin{eqnarray}
&& \frac{\partial}{\partial_{\alpha_i}} \left[ \int_{\Omega} \frac{ f^2(x) }{ g(x) } dx - \mu^2 + \lambda \alpha_i \right] = 0 \nonumber \\
&\Leftrightarrow&
\int_{\Omega} b_i(x) \frac{ f^2(x) }{ g^2(x) } dx = \lambda.
\end{eqnarray}
This tells us that for the optimal $g$, the projection of ${f^2}/{g^2}$ on the basis functions must be a constant, i.e:
\begin{equation}
<f^2/g^2, b_i>\,\, =\,\, <f^2/g^2, b_j> \quad \forall (i,j)
\end{equation}
which can also be expressed as the equivalent condition:
\begin{equation}
<f^2/g^2, b_i - b_j>\,\, =\,\, 0 \quad \forall (i,j)
\end{equation}
however, it doesn't yet provide a closed formula for finding proper coefficients $\alpha_i$ to realize it. Notice that one could also equivalently express (20) and (21) as:
\begin{eqnarray}
&& E_{b_i}[ f^2/g^2 ] = E_{b_j}[ f^2/g^2 ] \quad \forall (i,j) \nonumber \\
&\Leftrightarrow& E_{b_i - b_j}[ f^2/g^2 ] = 0 \quad \forall (i,j)
\end{eqnarray}
To gather further insights we reformulate the problem in a slightly different way. We can treat our function $g = \sum_i \alpha_i b_i(x)$ as a multiple importance sampling combination \cite{Veach:1995}, where $b_i$ represents the density of the $i$-th technique and the coefficients $\alpha_i$ represent the relative mixture weights, or sampling frequencies.
Under this light, the problem is essentially the same as that analysed by He and Owen \cite{He:2014}. Unfortunately, while He and Owen demonstrated that variance is jointly convex in the mixture weights, thus allowing to use a convex optimization algorithm to find an optimum, a closed form solution is yet to be found.

\ifnum 1 = 0
However, we can still use this basic multiple importance sampling framework to simplify the problem. Assume that sampling from $g$ is carried out by sampling from each component $b_i$ independently, and using the multiple importance sampling estimator:
\begin{equation}
F_N = \frac{1}{N} \sum_{i=1}^n \sum_{j=1}^{n_i} w_i(X_{i,j}) \frac{ f(X_{i,j}) }{ b_i(X_{i,j})}
\end{equation}
where $N = \sum_j n_j$, $n_i$ is proportional to $\alpha_i$, i.e. $n_i/N \approx \alpha_i$, and $w_i$ is the balance heuristic weight:
\begin{equation}
w_i(x) = \frac{b_i(x)}{\sum_j \alpha_j b_j(x)}
\end{equation}
The variance of this estimator can now be written as the sum of the variances of the individual components:
\begin{equation}
\sum_i \alpha_i \int_\Omega b_i(x) f^2(x) / g^2(x) dx
\end{equation}
\fi

\section{Optimal approximation of a mixture}

Now, let's go back to the second problem mentioned in the introduction.
Suppose we have a finite family of functions $\mathcal{F} = (f_i : \Omega \rightarrow \mathbb{R})_{i=1,\dots,n}$, and an equally sized family of importance sampling densities $\mathcal{G} = ( g_1, ..., g_n )$, each of which can be used to \emph{approximately} sample from the corresponding term in $\mathcal{F}$, i.e. such that $g_i \approx f_i$.
And let's further consider the problem of estimating the integral of a finite mixture:
\begin{equation}
f(x) = \sum_{i=1}^n \alpha_i f_i(x)
\end{equation}
sampling from the densities $g_i$ with frequency $\tilde{\alpha}_i$, and using the importance sampling estimator:
\begin{equation}
E[f] = \sum_{i=1}^n \tilde{\alpha}_i E_{g_i}
\left[
	\frac{\alpha_i}{\tilde{\alpha}_i} \cdot \frac{f_i}{g_i}
\right].
\end{equation}
which can be realized sampling $N = \sum_i N_i$ random variables $X_{i,j} \sim g_i$, with $N_i/N \approx \tilde{\alpha}_i$, and applying the summation rule:
\begin{equation}
	E[f] \approx \frac{1}{N} \sum_{i=1}^n \sum_{j=1}^{N_i} w_i(X_{i,j})
\end{equation}
with proper importance sampling weights:
\begin{equation}
w_i(x) = \frac{\alpha_i}{\tilde{\alpha}_i} \cdot \frac{f_i(x)}{g_i(x)}.
\end{equation}
Normally, one could consider setting $\tilde{\alpha}_i = \alpha_i$, but we can again ask ourselves whether this is indeed an optimal choice.
If we consider the resulting sample weights as realizations of a random variable $W$,
we can now write the variance of the estimator as:
\begin{equation}
V = E[W^2] - E[W]^2.
\end{equation}
with the first term being equal to:
\begin{eqnarray}
E[W^2] &=& \sum_i \tilde{\alpha}_i \int_\Omega w_i^2(x) g_i(x) dx \nonumber \\
&=& \sum_i \frac{\alpha_i^2}{\tilde{\alpha}_i} \int_\Omega \frac{f_i^2(x)}{g_i^2(x)} g_i(x) dx \nonumber \\
&=& \sum_i \frac{\alpha_i^2}{\tilde{\alpha}_i} \int_\Omega \frac{f_i^2(x)}{g_i(x)} dx
\end{eqnarray}
and the second term equal to the squared integral of $f$:
\begin{eqnarray}
E[W]^2 &=& \left(\sum_i \tilde{\alpha}_i \int_\Omega w_i(x) g_i(x) dx\right)^2 \nonumber \\
&=& \left(\sum_i \alpha_i \int_\Omega f_i(x) dx\right)^2 \nonumber \\
&=& \mu^2
\end{eqnarray}
Hence, we obtain:
\begin{equation}
V = \sum_i \frac{\alpha_i^2}{\tilde{\alpha}_i} \int_\Omega \frac{f_i^2(x)}{g_i(x)} dx - \mu^2
\end{equation}
Now, posing:
\begin{equation}
m_i^{(2)} = \alpha_i^2 \int_\Omega \frac{ f_i^2(x) }{g_i(x)} dx \nonumber
\end{equation}
and applying again the method of Lagrange multipliers to solve for optimal values of $\tilde{\alpha}_i$ with the normalization constraint $\sum_i \tilde{\alpha}_i = 1$, we obtain:
\begin{equation}
\tilde{\alpha}_i =
\frac{
	\sqrt{m_i^{(2)}}
}{
	\sum_j \sqrt{m_j^{(2)}}
}.
\end{equation}
Notice the similarity with equation (18) from the previous section.
As mentioned in the introduction, this similarity is not a coincidence, as the result of equation (18) can be seen as a particular case of this derivation, by setting $b_i = g_i$, and considering each $f_i$ as the restriction of $f$ to the non-overlapping supports $C_i$, i.e. $f_i = f_{|C_i}$.
However, as we will shortly see, it is important to make some careful considerations.

\vspace{3mm}
\subsection{Generalization to multiple importance sampling estimators}

In this section, we have assumed a partitioning of $f$ into a weighted sum of functions $f_i$ that can be each \emph{individually} sampled by corresponding densities $g_i$.
However, such a strict partitioning might not always be desirable, or lead to an optimal estimator.
In fact, if the functions $f_i$ have overlapping supports, and the densities $g_i$ are only locally good approximations for \emph{some} of the features of $f_i$, a better estimator might be obtained considering a multiple importance sampling combination.
For example, we could use the unbiased estimator:
\begin{equation}
E[f] = \sum_{i=1}^n \tilde{\alpha}_i E_{g_i}
\left[
	\frac{f}
	{\sum_i \tilde{\alpha}_i g_i}
\right]
\end{equation}
which is obtained applying the balance heuristic \cite{Veach:1995}, and resulting in the alternative sample weights:
\begin{equation}
w(x) = \frac{ f(x) }{\sum_j \tilde{\alpha}_j g_j(x)}
\end{equation}
Optimizing the variance of such an estimator would now become exactly the same problem as finding the optimal projection of $f$ onto a set of arbitrary basis functions, as seen in section 2.2 and as analysed by He and Owen \shortcite{He:2014}.

\section{Discussion}

In this work we have provided simple closed form formulas to build optimal importance sampling approximations within certain classes of mixture distributions, and showed how the resulting projection coefficients differ from the commonly used $\ell_1$ projections.
We believe our results to be potentially useful for the solution of many Monte Carlo integration problems.
Particularly, there could be numerous applications for path sampling in the field of light transport simulation, ranging from the construction of better radiance or importance field approximations for path guiding \cite{Vorba:2014:OLP,Dahm:2017}, to improved importance caching distributions \cite{Georgiev:2012:IC}, to new methods for sampling direct illumination.

\paragraph*{Acknowledgements:}
We thank Peter Shirley for suggesting the use of Lagrange multipliers and providing useful feedback on the early drafts of this paper.

\bibliographystyle{acmsiggraph}
\bibliography{main}

\end{document}